\documentclass[prl,twocolumn,aps,superscriptaddress,10pt,floatfix]{revtex4-2}

\usepackage{amsmath}
\usepackage{amssymb}
\usepackage{txfonts}
\usepackage{siunitx}
\usepackage{graphicx}
\usepackage{bm}
\usepackage{hyperref}
\usepackage{float}
\usepackage{braket}
\usepackage[T1]{fontenc}
\usepackage[latin1]{inputenc}
\usepackage[table]{xcolor}
\usepackage{lipsum}     
\usepackage[most]{tcolorbox}
\usepackage{blindtext}
\hypersetup{
	colorlinks=true,
	citecolor=blue,
	linkcolor=blue,
	urlcolor=blue}

\newcommand{\refeq}[1]{Eq.~(\ref{#1})}

\newcommand{\reffig}[1]{Fig.~\ref{#1}}

\newcommand{\punc}[1]{\,{\text{#1}}}
\newcommand{\sub}[1]{_{\text{#1}}}

\newcommand{\super}[1]{^{\text{#1}}}
\newcommand{\Chi}{\mathrm{X}}

\newcommand{\Wm}{\mathbf{W}}
\newcommand{\deltav}{\boldsymbol{\delta}}

\DeclareMathOperator{\Res}{Res}

\DeclareMathOperator{\erf}{erf}
\DeclareMathOperator{\erfc}{erfc}

\newcommand{\DTO}{Dy$_2$Ti$_2$O$_7$}

\begin{document}

\title{Dynamic scaling theory for a field quench near the Kasteleyn transition in spin ice} 
\author{Stephen Powell}
\affiliation{School of Physics and Astronomy, The University of Nottingham, Nottingham, NG7 2RD, United Kingdom}
\author{Sukla Pal}
\affiliation{School of Physics and Astronomy, The University of Nottingham, Nottingham, NG7 2RD, United Kingdom}
\affiliation{Pitaevskii BEC Center, CNR-INO and Dipartimento di Fisica, Universit\`a di Trento, I-38123 Trento, Italy}

\begin{abstract}
We present a dynamic scaling theory to describe relaxation dynamics following a magnetic-field quench near an unconventional phase transition in the magnetic material spin ice. Starting from a microscopic model, we derive an effective description for the critical dynamics in terms of the seeding and growth of string excitations, and use this to find scaling forms in terms of time, reduced temperature and monopole fugacity. We confirm the predictions of scaling theory using Monte Carlo simulations, which also show good quantitative agreement with analytical expressions valid in the limit of low monopole density. As well as being relevant for experiments in the spin ice materials, our results open the way for the study of dynamic critical properties in a family of unconventional classical phase transitions.
\end{abstract}

\maketitle

Phase transitions that go beyond the ``Landau paradigm'', the standard framework for critical phenomena, are those where critical properties cannot be described purely in terms of long-wavelength modes of an order parameter \cite{Senthil2024}. A number of examples have been proposed in quantum models where apparently unrelated order parameters are connected by the nature of their topological defects \cite{Senthil2004a,Senthil2004b,Balents2005,Nussinov2007}.

Non-Landau transitions also occur in classical systems with strong constraints \cite{Alet2006,Jaubert2008,Chen2009,Papanikolaou2010,Wilkins2019,Wilkins2020}, and can be described either in terms of ``topological order'' \cite{Castelnovo2012} or through the confinement of defects that act as charges in an emergent gauge theory \cite{Henley2010}. Transitions of this type have been identified in dimer models \cite{Henley2010} and in models of geometrically frustrated magnets \cite{Balents2010}. The latter include a class of materials known as spin ice \cite{Harris1997,Bramwell2020}, where there is an extensive set of approximately degenerate ground states, the excitations above which are magnetic monopoles \cite{Castelnovo2008}. It is by now well understood how to describe \emph{static} critical properties at these transitions, including how to incorporate monopoles \cite{Powell2012,Powell2013}, in terms of appropriate critical fields.

In this work, we extend the study of this family of unconventional phase transitions to dynamic critical properties \cite{Hohenberg1977}. By applying general principles of scaling and the renormalization group to time-dependent phenomena, scaling theory can describe dynamics at and near critical points \cite{Goldenfeld,Liu2014}, and has recently been applied to topological phases in a range of classical systems \cite{Jelic2011,Hamp2015,Xu2018}. Here, we demonstrate how it can be applied to unconventional classical criticality.

We start from a microscopic model of spin ice that has a phase transition at a critical strength of the applied magnetic field \cite{Jaubert2008,Powell2008}, and derive dynamic scaling forms for the relaxation of the magnetization and monopole density following a quench of the field to near its critical value. We also present results of Monte Carlo simulations, which confirm the scaling forms and show close quantitative agreement (with no fitting parameters) with exact analytical expressions for the scaling functions that we derive in the limit of low monopole density.

This is the first such theory of dynamic critical phenomena at an unconventional phase transition of this type and opens the way to understanding dynamics near unconventional transitions more broadly. Our results are also of direct relevance for experiments in the spin ice materials, with scaling theory putting strong constraints on the magnetization relaxation near the transition, which is measurable in experiments \cite{Mostame2014b}. Equilibration is hindered at low temperature in spin ice \cite{Giblin2018}, making it difficult to compare with scaling theory for static critical properties; its extension into the dynamics potentially allows for the direct experimental observation of unconventional critical properties.

\begin{figure}[b]
\includegraphics[width=\columnwidth]{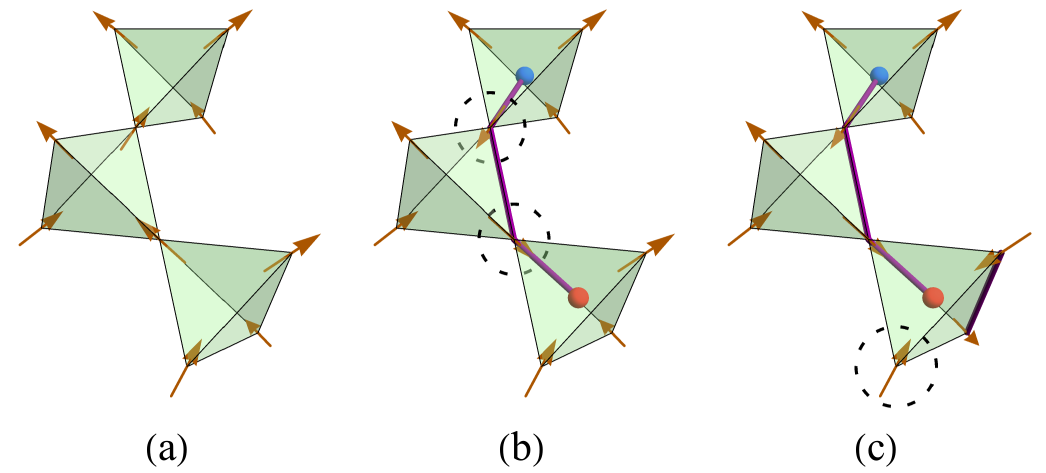}
\caption{Illustration of string growth. (a) In the starting configuration, all spins point upwards. (b) Flipping two spins (circled) gives a pair of monopoles joined by a string of length \(\ell=2\). The string could be extended to \(\ell=3\) by flipping either of the two bottom-most spins. (c) When a second string passes through the bottom tetrahedron, only one spin remains (circled) that can be flipped to extend the string.}
\label{FigString}
\end{figure}

\textit{Model}---We start from a minimal model of spin ice, with classical spins \(\boldsymbol{S}_i=\sigma_i\hat{\boldsymbol{n}}_i\) on the sites \(i\) of a pyrochlore lattice, a network of corner-sharing tetrahedra (see Figure~\ref{FigString}). The fixed unit vector \(\hat{\boldsymbol{n}}_i\) points along the local \(\langle 111 \rangle\) easy axis between the centers of the two tetrahedra to which site \(i\) belongs, and \(\sigma_i = \pm 1\) is an effective Ising degree of freedom \cite{Bramwell2020}. We include only nearest-neighbor interactions, and write the Hamiltonian as \(H = \Delta\sum_\alpha n_\alpha^2\), where
\(n_\alpha = -\frac{1}{2}\eta_\alpha \sum_{i \in \alpha} \sigma_i\), which takes integer values,
is the ``magnetic charge'' on tetrahedron \(\alpha\) (\(\eta_\alpha = \pm1\) for up- and down-pointing tetrahedra) \cite{Pal2024}. There is an extensive set of ground states with \(H=0\), in which every tetrahedron has two spins pointing in and two pointing out; the elementary excitations are ``monopoles'' with \(n_\alpha = \pm1\) and energy \( \Delta = \frac{2}{3}J + \frac{8}{3}\left(1 + \sqrt{\frac{2}{3}}\right)D\) \cite{Castelnovo2008}, in terms of the exchange \(J\) and dipolar \(D\) coupling constants. In our numerical results, we set \(\Delta = \qty{2.8}{\kelvin}\), corresponding to \DTO\ \cite{Gardner2010}.

An external magnetic field \(\boldsymbol{h}\) applied along the \(z\) direction couples to the spins through a term \(H_h = -\mu \sum_i \boldsymbol{h}\cdot\boldsymbol{S}_i\),
where \(\mu \simeq 10\mu\sub{B}\) is the magnetic moment of each spin \cite{Gardner2010}. We define \(h = \mu\lvert \boldsymbol{h}\rvert/\sqrt{3}\) and
\begin{equation}
m_z = \frac{\sqrt{3}}{N\sub{s}}\sum_i \left(\boldsymbol{S}_i\right)_z\punc,
\end{equation}
where \(N\sub{s}\) is the number of spins. The normalization is chosen so that the maximum value is \(m_z = 1\) and \(H_h/N\sub{s} = -m_z h\).

In equilibrium, the system has a Kasteleyn transition \cite{Kasteleyn1963}, in the limit of zero monopole fugacity \(z=e^{-\Delta/T}\), at a temperature \(T = T\sub{K} \equiv \frac{2}{\ln 2}h\) \cite{Jaubert2008}. (We set \(k\sub{B}=1\).) The magnetization is saturated, \(m_z=1\), for \(T < T\sub{K}\), and the deviation from this value, \(\sigma = 1 - {m_z}\), increases continuously for \(T > T\sub{K}\), as strings of flipped spins, as illustrated in \reffig{FigString}, appear with finite density. For \(z>0\), the transition is replaced by a crossover \cite{Jaubert2008}, which can be understood using static scaling theory \cite{Powell2013}.

We consider a quench starting from \(h=+\infty\), deep within the saturated phase, to a final value \(h\) such that \(T \simeq T\sub{K}\). We use a minimal model for the dynamics: spin flips are attempted at rate \(N\sub{s}\), with time \(t\) measured in units of \(\tau\sub{flip}\simeq \qty{3}{\ms}\) \cite{Ryzhkin2005,Jaubert2009}, and accepted with Glauber probability \(P\sub{G}(\delta E) = (e^{\delta E/T} + 1)^{-1}\) \cite{Glauber1963,Suzen2014}, where \(\delta E\) is the energy change.

\textit{Single-string model}---At the equilibrium Kasteleyn transition, the relevant scaling variables are \(z\) and reduced temperature \(\theta = (T - T\sub{K})/{T\sub{K}}\) \cite{Powell2013}. We therefore expect the full scaling behavior of the dynamics to involve these two variables as well as time \(t\). As a first step, we employ a single-string model \cite{Pal2024}, which, we will argue, applies in the limit of vanishing \(z\). Using this, we calculate dynamic scaling forms for the magnetization and monopole density that are valid for long time and temperature \(T\) close to \(T\sub{K}\).

Consider a stochastic process, with rate matrix \(\Wm\), describing a population of isolated strings, with \(N(\ell, t)\) giving the number of strings of length \(\ell \ge 1\) at time \(t\) after the quench. The evolution of the string population is governed by four processes: (1) A string can be created by flipping any spin downwards, creating a pair of monopoles joined by a string of length \(\ell = 1\). This occurs at a rate \(N\sub{s}r_*\) with \(r_* = P\sub{G}(2\Delta + 2h)=(e^{2(\Delta + h)/T} + 1)^{-1}\), where the energy change includes contributions \(\Delta\) for each monopole plus the change in \(H_h\) due to the (single) flipped spin.

(2) The string can be extended by moving one of the two monopoles and thereby increasing their separation; this requires flipping one of \(4\) possible upward-pointing spins (two for each monopole). This increases the energy by \(2 h\) and therefore occurs with rate
\(W_{\ell+1,\ell} = r_+ = 4(e^{2h/T} + 1)^{-1}\). (3) Shrinking the string requires flipping one of \(2\) downward-pointing spins (one for each monopole), so decreases the energy by the same amount, and hence has rate
\(W_{\ell,\ell+1} = r_- = 2(e^{-2h/T} + 1)^{-1}\). (4) Removing a string with \(\ell=1\) requires flipping the one remaining downward-pointing spin. It changes the energy by \(\delta E = -2\Delta - 2h\) and so has rate
\(r_0 = (e^{-2(\Delta + h)/T} + 1)^{-1}\).

The rates of growth and shrinkage are equal, \(r_+ = r_-\), when \(e^{2h/T} = 2\), which occurs at the transition temperature, \(T = T\sub{K}\). Near the critical point, i.e., for \(T\simeq T\sub{K}\) and small \(z\), \(r_* = \frac{1}{2}z^2 + O(z^4, \theta z^2)\), \(r_0 = 1 + O(z^2)\), and
\(
{r_+}/{r_-} = 1 + \theta \ln 2 + O(\theta^2)\).

Treating all strings as isolated, the total number \(n_1\) of monopoles is double the number of strings. Their density is therefore \(\rho_1 = n_1/N\sub{d} = 4r_*M_0(t)\), where \(N\sub{d} = \frac{1}{2}N\sub{s}\) is the number of tetrahedra and
\begin{equation}
\label{eq:DefineMnt}
M_n(t) = \frac{1}{N\sub{s}r_*}\sum_{\ell = 1}^{\infty} \ell^n N(\ell,t)\punc.
\end{equation}
A string of length \(\ell\) reduces \(m_z\) by \(2\ell\), and so
\(\sigma = 2r_*M_1(t)\), proportional to the mean areal density of strings.

The dynamics starts at \(t=0\) from the saturated configuration, and so \(N(\ell,t)\) obeys the initial condition \(N(\ell, 0) = 0\) for all \(\ell \ge 1\). Taking into account the four processes described above, it evolves according to
\begin{equation}
\label{eq:NelltODE}
\frac{\partial}{\partial t}N(\ell, t) = \delta_{\ell,1}N\sub{s}r_* + \sum_{\ell'=1}^{\infty} W_{\ell,\ell'}N(\ell',t)
\punc,
\end{equation}
where \(W_{1,1} = -(r_0+r_+)\) and \(W_{\ell,\ell} = -(r_-+r_+)\) for \(\ell>1\). We first solve this model exactly, finding the leading-order behavior at long time near the critical point, before showing how the same results can be derived using a continuum-\(\ell\) description.

The solution of \refeq{eq:NelltODE} is
\begin{equation}
\label{eq:Nellt}
N(\ell, t) = N\sub{s}r_* \int_0^t dt' \,\deltav_\ell\cdot e^{\Wm t'}\deltav_1 \punc,
\end{equation}
where \(\deltav_\ell\) is a unit vector. The eigenvectors of \(\Wm\) are linear combinations of right- and leftward moving waves, \(q^\ell\) and \(q^{-\ell}\), where \(q\) is a complex number, with eigenvalues \(\lambda(q) = r_+(q^{-1}-1) + r_-(q - 1)\). Writing \(\deltav_1\) in terms of them  \cite{Pal2024} and performing the integral in \refeq{eq:Nellt} gives
\(M_n(t) = (2\pi i)^{-1}\oint_{\mathfrak{C}} d q \left[f_n(q) - \left. f_n(q)\right\rvert_{t=0}\right]\), where
\begin{equation}
f_n(q) = \frac{1}{q} \frac{1}{(1-q)^{n+1}}\frac{r_+ - r_- q^2}{r_+ + (r_0 - r_-)q}\frac{e^{t\lambda(q)}}{\lambda(q)}
\end{equation}
and the contour \(\mathfrak{C}\) is a counterclockwise circle of radius \(\le 1\). The second term in the integrand is a rational function of \(q\); its integral is \(\Res(f_n,r_+/r_-)\) for \(\theta < 0\) and zero otherwise, where \(\Res\) denotes the residue.

For \(t\gg \tau\equiv(r_+r_-)^{-1/2}\), we can evaluate the contour integral by deforming \(\mathfrak{C}\) to pass through the saddle point, \(\lambda'(q_0)=0\), at \(q_0 = \sqrt{r_+/r_-}\). For \(\theta > 0\), this involves crossing the pole at \(q=1\), adding a contribution given by its residue. One can therefore write
\begin{equation}
\label{eq:MntIntegral}
M_n(t) = \oint_{\mathfrak{C}'} \frac{d q}{2\pi i} f_n(q) - \Theta(\theta)\Res(f_n,1) - \Theta(-\theta)\Res(f_n,q_0^2)
\punc,
\end{equation}
where \(\mathfrak{C}'\) is a circle of radius \(q_0\) and \(\Theta\) is the unit step function.

Expanding near the saddle point gives \(t\lambda(q_0 + i \xi\sqrt{\tau/t})\approx -x^2 - \xi^2\) to leading order in \(\theta\), where \(x = \frac{\ln 2}{\sqrt{3}}\theta t^{1/2}\). The integral is therefore dominated by \(\xi = O(1)\) and can be extended to \(-\infty < \xi <\infty\). For small \(\theta\), the poles \(q_+=1\) and \(q_-=q_0^2\) also approach the saddle point, and are given by \(\xi_\pm \approx \pm i x\) to the same order. The result is
\begin{equation}
M_n(t) \approx \left(\frac{4t}{3}\right)^{(n+1)/2}\left[\int_{-\infty}^{\infty}\frac{d\xi}{2\pi}g_n(\xi) - i \Res(g_n,i\lvert x \rvert)\right]
\punc,
\end{equation}
where
\(g_n(\xi) = 2i \xi e^{-\xi^2-x^2}(-x-i\xi)^{-n-1}(x^2+\xi^2)^{-1}\)
is found by expanding \(f_n(q_0 + i \xi\sqrt{\tau/t})\) for large \(t\) and small \(\lvert \theta\rvert\).

The integral over \(\xi\) can be expressed in terms of the error function \(\erf x\), giving \(M_0(t) \approx \frac{1}{2}t^{1/2}\Psi(\theta t^{1/2})\) and \(M_1(t) \approx t\Phi(\theta t^{1/2})\), where
\begin{align}\label{eq:Psi}
\Psi\left(\!\frac{\sqrt{3}x}{\ln 2}\!\right) &= \frac{4}{\sqrt{3}}\left[x + \frac{e^{-x^2}}{\sqrt{\pi}} + \left(x + \frac{1}{2x}\right)\erf x\right]\\
\label{eq:Phi}
\Phi\left(\!\frac{\sqrt{3}x}{\ln 2}\!\right) &= \frac{4}{3}\left[1 + x^2 + \frac{e^{-x^2}}{\sqrt{\pi}}\left(x + \frac{1}{2x}\right) + \left(1 - \frac{1}{4x^2}+x^2\right)\erf x\right]
\end{align}
are defined such that
\begin{equation}
\begin{aligned}
\rho_1 &\approx z^2 t^{1/2}\Psi(\theta t^{1/2})&\quad&\quad&
\sigma &\approx z^2 t\Phi(\theta t^{1/2})\punc.
\end{aligned}
\label{eq:ScalingForms1}
\end{equation}
These are dynamic scaling forms for the monopole density and magnetization, depending on \(\theta\) only through the combination \(\theta t^{1/2}\) for long times close to the critical point.

The same scaling results can be found starting from \refeq{eq:NelltODE} by using a continuum-\(\ell\) description. This is justified at long times near the critical point because the integral giving \(M_n(t)\), \refeq{eq:MntIntegral}, is then dominated by a small region near \(q=1\), where the eigenvectors are slowly varying functions of \(\ell\). We therefore write \(N(\ell,t)\approx N\sub{s}\varphi(\ell,t)\) for \(\ell \ge 1\) where \(\varphi\) is, by assumption, a smooth function of \(\ell\) and \(t\). Expanding \refeq{eq:NelltODE} and keeping derivatives up to second order in \(\ell\) gives
\begin{equation}
\label{eq:NelltPDE}
\frac{\partial}{\partial t}\varphi(\ell, t) = -(r_+ - r_-) \frac{\partial}{\partial \ell}\varphi(\ell, t) + \frac{1}{2}(r_++r_-)\frac{\partial^2}{\partial\ell^2}\varphi(\ell,t)\punc,
\end{equation}
with the boundary condition \(\varphi(0, t) = r_*/(r_0 + r_+ - r_-)\). This describes a diffusion process with a source at \(\ell = 0\) and a drift term whose coefficient \(r_+ - r_-\) changes sign at \(T\sub{K}\).

Near the critical point, expanding to leading order in \(\theta\) and \(z\) gives \(r_+-r_-\approx \frac{4}{3}\theta \ln 2\), \(r_++r_- \approx \frac{8}{3}\), and \(\varphi(0,t) = \frac{1}{2}z^2\). For any initial distribution peaked near \(\ell = 0\), the solution at long times is insensitive to initial conditions and can be expressed as \(\varphi(\ell, t) = z^2\Chi(\ell t^{-1/2}, \theta t^{1/2})\), with
\begin{equation}
\Chi\left(\!\frac{4y}{\sqrt{3}},\frac{\sqrt{3}x}{\ln 2}\!\right) = \frac{1}{4}\left[\erfc(y-x)+ e^{4xy}\erfc(y+x)\right]\punc,
\end{equation}
where \(\erfc x = 1 - \erf x\) is the complementary error function. Using \refeq{eq:DefineMnt}, with the sum over \(\ell\) replaced by an integral, then reproduces the scaling forms of \refeq{eq:ScalingForms1} with the same functions \(\Psi\) and \(\Phi\).

\textit{Generalized scaling}---The single-string model assumes that strings are sparse enough to neglect interactions between them. Since new strings appear at a rate \(\propto r_* \sim z^2\), it applies only the limit of vanishing \(z\). Including \(z\) within the scaling form therefore requires going beyond the single-string picture.
  
Consider the situation where one of the two monopoles at either end of a string occupies a tetrahedron that another string passes through, illustrated in \reffig{FigString}(c). (In other words, another two spins are flipped downwards on the tetrahedron.) In this case, only one up spin remains that can be flipped to extend the string, and so the rate for this process is decreased from \(r_+\). In addition, there are two down spins, rather than one, that can be flipped back up, and so the rate of shrinking the string is greater than \(r_-\). This excluded-volume effect is equivalent to the hard-core repulsion between strings that reduces their entropy at finite density and allows for a liquid of strings in equilibrium \cite{Jaubert2008}.
  
To include this effect within a mean-field approximation, we reduce the drift coefficient \(r_+ - r_-\) in \refeq{eq:NelltPDE} by an amount proportional to the areal string density \(\sigma\). To leading order in \(\theta\), this gives
\begin{equation}
\label{eq:meanfieldPDE}
\frac{\partial\varphi}{\partial t} = -\frac{4\ln2}{3}\left[\theta - b\int_0^\infty d\ell\,\ell \varphi(\ell,t) \right] \frac{\partial\varphi}{\partial \ell} + \frac{4}{3}\frac{\partial^2\varphi}{\partial\ell^2}\punc,
\end{equation}
where \(b\) is a positive constant independent of \(\theta\) and \(z\). By rescaling the parameters, one can write the long-time solution of this nonlinear integro-differential equation as \(\varphi(\ell, t) = z^2\Chi(\ell t^{-1/2}, \theta t^{1/2}, z\lvert\theta\rvert^{-3/2})\). Integrating over \(\ell\) then gives the generalized scaling behavior
\begin{equation}
\begin{aligned}
\rho_1 &\approx z^2 t^{1/2} \Psi(\theta t^{1/2},z\lvert\theta\rvert^{-3/2})\\
\sigma &\approx z^2 t \Phi(\theta t^{1/2},z\lvert\theta\rvert^{-3/2})\punc,
\end{aligned}
\label{eq:ScalingForms2}
\end{equation}
describing the dependence on \(t\), \(\theta\), and \(z\). The two-parameter functions \(\Psi\) and \(\Phi\) are determined by the solution of \refeq{eq:meanfieldPDE} and depend on the phenomenological parameter \(b\). For \(z\rightarrow 0\), \refeq{eq:ScalingForms2} should reduce to \refeq{eq:ScalingForms1}, meaning that \(\Psi(u,v)\rightarrow \Psi(u)\) and \(\Phi(u,v)\rightarrow\Phi(u)\) for \(v\rightarrow0\).
  
In the long-time limit, \(t\rightarrow\infty\), they should instead reduce to the equilibrium scaling forms \cite{Powell2013}, which can be expressed as \(\rho_1 \approx \lvert\theta\rvert^{2}\Psi\super{eq}_\pm(z\lvert\theta\rvert^{-3/2})\) and \(\sigma \approx \lvert\theta\rvert\Phi\super{eq}_\pm(z\lvert\theta\rvert^{-3/2})\) \footnote{The scaling functions, calculated using mean-field theory \cite{Jaubert2008,Powell2013}, are \(\Psi_\pm\super{eq}(v)=4v^{4/3}\psi\left(\pm \frac{2\ln2}{3} v^{-2/3}\right)\) and \(\Phi_\pm\super{eq}(v) = 4v^{2/3}\left[\psi\left(\pm \frac{2\ln2}{3} v^{-2/3}\right)\right]^2\), where \(\psi(x)\) is the positive solution of \(\psi^3-x \psi - \frac{1}{3}=0\).}, up to logarithmic corrections. (The subscript \(\pm\) indicates that the functions also depend on the sign of \(\theta\). The rational exponents and logarithmic corrections are due to the fact that the equilibrium transition is at its upper critical dimension \cite{Jaubert2008}.) Our generalized scaling forms are indeed compatible with these equilibrium results, despite being derived using mean-field theories that are formulated quite differently. By matching the functions, we infer the limits \(\Psi(u,v)\rightarrow \lvert u\rvert^{-1}v^{-2}\Psi\super{eq}_\pm(v)\) and \(\Phi(u,v)\rightarrow u^{-2}v^{-2}\Phi\super{eq}_\pm(v)\) for \(u\rightarrow\pm\infty\).

\textit{Results}---We first compare Monte Carlo (MC) simulations to the single-string model, \refeq{eq:ScalingForms1}, in terms of which nonzero \(z\) gives corrections to scaling. For fixed \(z\), we expect agreement for \(t\) large enough that scaling applies but not so large that the corrections become significant \cite{Goldenfeld}; the general scaling forms of \refeq{eq:ScalingForms2} imply that the latter occurs for \(t \sim z^{-4/3}\). We similarly require small \(\lvert \theta\rvert\), but above a value \(\lvert\theta\rvert\sim z^{2/3}\).

\begin{figure}
\centering
\includegraphics[width=0.925\linewidth]{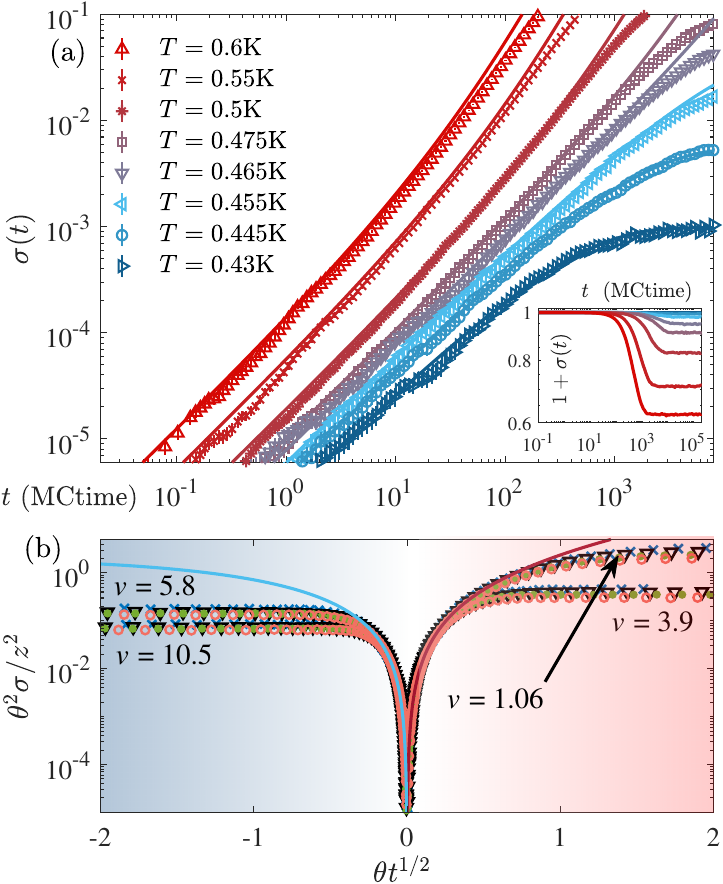}
\caption{(a) Monte Carlo results (symbols) for deviation from saturation magnetization, \(\sigma = 1 - m_z\), versus time \(t\) after the quench, with field \(h=\qty{0.16}{\kelvin}\), for which the transition temperature is \(T\sub{K} = \qty{0.462}{\kelvin}\), and a lattice of \(N\sub{s} = 128000\) spins. For each temperature \(T\), the solid line shows the analytical expression, \refeq{eq:ScalingForms1}, for the limit of low monopole fugacity \(z\). Error bars are shown but are mostly smaller than the symbols. (b) Results for \(h=\qty{0.16}{\kelvin}\) (blue cross), \(\qty{0.21}{\kelvin}\) (black triangle), \(\qty{0.25}{\kelvin}\) (green filled circle), and \(\qty{0.3}{\kelvin}\) (orange empty circle), with \(T\) chosen for each to give the same four values of \(v = z\lvert\theta\rvert^{-3/2}\), as labeled. (Error bars are roughly the symbol sizes.) As predicted by the generalized scaling forms, \refeq{eq:ScalingForms2}, when plotted as a function of \(u = \theta t^{1/2}\), data for the same \(v\) collapse onto a single curve. Thin solid lines show the analytical result for \(v\rightarrow 0\).}
\label{fig:MzScaling}
\end{figure} 
  
\begin{figure}
\centering
\includegraphics[width=0.925\linewidth]{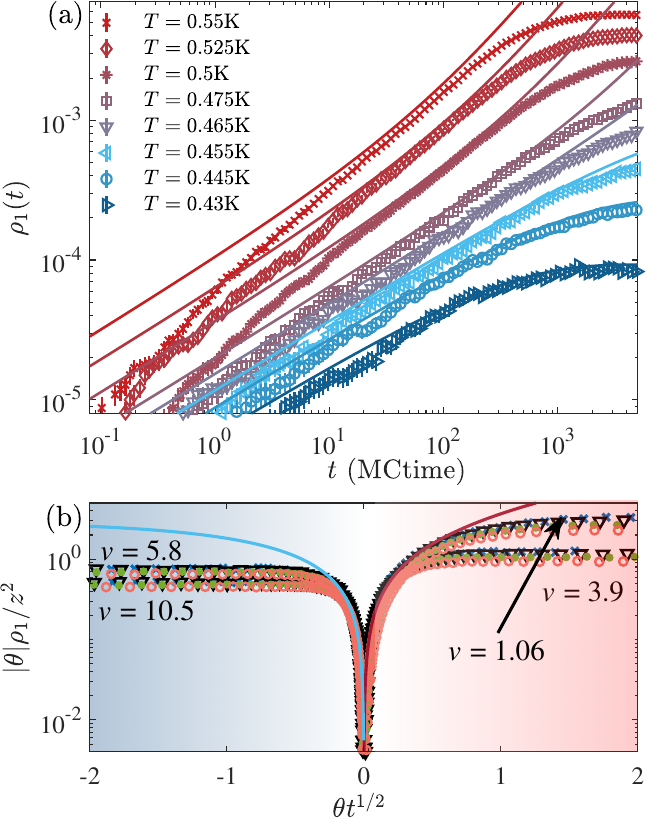}
\caption{(a) Monte Carlo results (symbols) for monopole density, \(\rho_1\), versus time \(t\) after the quench, for \(h=\qty{0.16}{\kelvin}\). The solid lines show the analytical expression, \refeq{eq:ScalingForms1}, for the limit of low monopole fugacity \(z\). (b) Results for various \(h\) plotted as a function of \(u = \theta t^{1/2}\), as in \reffig{fig:MzScaling}(b).}
\label{fig:MpoleScaling}
\end{figure}

Figures~\ref{fig:MzScaling}(a) and \ref{fig:MpoleScaling}(a) show MC results for \(h=\qty{0.16}{\kelvin}\) compared with \refeq{eq:ScalingForms1}. For \(T\) either close to or below \(T\sub{K} = \qty{0.462}{\kelvin}\), we see close quantitative agreement for \(\sigma\) over several decades of \(t\), with no fitting parameters. Agreement with \(\rho_1\) is also seen, though over a narrower region of \(t\) and \(T\), especially above \(T\sub{K}\). For larger values of \(h\) (not shown), deviations become progressively larger, as expected because \(T\), and hence \(z\), is then larger at the critical temperature.

The fact that \(\sigma\) follows the scaling form more closely than \(\rho_1\) is a consequence of using the large-\(t'\) behavior of \(e^{\Wm t'}\) in \refeq{eq:Nellt}. This assumes that the (finite) contribution from small \(t'\) (i.e., recently created strings) is negligible, which is true as long as the rest of the integral grows without bound as \(t\) gets larger. The mean string lifetime grows at least as \(\sim \lvert \theta \rvert^{-1}\), and so this assumption is valid for small \(\theta\) even for \(n=0\) (i.e., for \(\rho_1\)), but for \(n=1\) (\(\sigma\)) the contribution of longer (and hence typically older) strings has more weight in the sum.

Figures~\ref{fig:MzScaling}(b) and \ref{fig:MpoleScaling}(b) show scaled data for several values of \(h\), up to \(\qty{0.3}{\kelvin}\), with \(T\) chosen for each to give the same four values of \(v=z\lvert\theta\rvert^{-3/2}\). For both, data with the same \(v\) collapse onto the same function of \(u=\theta t^{1/2}\), in agreement with the generalized dynamic scaling forms, \refeq{eq:ScalingForms2}. The best collapse is again seen for \(\sigma\) and for smaller \(h\).

\textit{Conclusions}---Starting from a microscopic model of spin ice, we have derived dynamic scaling forms for relaxation of magnetization and monopole density following a magnetic field quench, in terms of time, reduced temperature, and monopole fugacity. We have also found analytical expressions for the scaling functions in the limit of low monopole fugacity, by solving an effective model for single strings. In both cases, we find close agreement with Monte Carlo simulations, demonstrating that the scaling forms provide a good description of the critical behavior for realistic parameter values. This may be particularly important for experiments in the classical spin ice materials \cite{Paulsen2014}, because of well-known issues reaching equilibrium \cite{Giblin2018}. The theory is potentially also relevant for quenches in artificial spin ice \cite{Libal2020}.

Our results are consistent with previous analytical studies of equilibrium critical properties at the same transition \cite{Powell2013}, which they extend to the dynamics. Using the dependence of the scaling forms, \refeq{eq:ScalingForms2}, on \(u=\theta t^{1/2}\) we infer \(\nu \mathfrak{z} = 2\), where \(\nu=\frac{1}{2}\) is the correlation-length exponent \cite{Powell2013} and \(\mathfrak{z}\) the dynamic critical exponent \cite{Hohenberg1977}, and hence \(\mathfrak{z} = 4\). This value is implied by the scaling forms in the single-string limit, \refeq{eq:ScalingForms1}, and so does not rely on the mean-field treatment used for generalized scaling.

We expect similar dynamic scaling behavior to occur at other related unconventional classical phase transitions. Note that the dynamic critical theory, \refeq{eq:meanfieldPDE}, neglects the spatial structure of the strings and their relative positions, and so is a mean-field theory in the spirit of the well-mixed approximation for chemical reactions \cite{Schnoerr2017}. A particularly interesting aspect from a theoretical point of view is that it is expressed in terms of a function \(\varphi\) of both time and an additional variable that corresponds to the length of a string. This echoes the critical theory for the equilibrium properties, where the direction of string growth acts as a fictitious ``time'' dimension \cite{Powell2008}.

\textit{Acknowledgments}---This work was supported by the Engineering and Physical Sciences Research Council grant number EP/T021691/1. The numerical simulations used resources provided by the University of Nottingham's Ada HPC service.

\textit{Research data statement}---Research data are available from the Nottingham Resarch Data Management Repository \cite{DataRepository7562}.

\bibliography{spinice}{}

%
%
 
\end{document}